\documentclass[pre,amssymb,reprint,superscriptaddress,showpacs]{revtex4-1}
\usepackage[dvipdfmx]{graphicx}
\usepackage{color}
\usepackage{ulem}

\newcommand{\rNA}[0]{{\bf r}^{\mbox{\scriptsize (NA)}}}
\newcommand{\gdot}[0]{\dot{\gamma}}

\renewcommand{\phi}{\varphi}
\newcommand{\phiJ}{\varphi_{\rm J}}

\newcommand{\odif}[2]{\frac{{\rm d} #1}{{\rm d} #2}}

\newcommand{\be}{\begin{equation}}
\newcommand{\ee}{\end{equation}}
\newcommand{\bea}{\begin{equnaray}}
\newcommand{\eea}{\end{equnaray}}
\newcommand{\ba}{\begin{align}}
\newcommand{\ea}{\end{align}}
\newcommand{\ave}[1]{\left\langle {#1} \right\rangle}

\begin{document}

\title{Classification of the reversible-irreversible transitions in particle trajectories across the jamming transition point}

\author{Kentaro Nagasawa}
\affiliation{Department of Physics, Nagoya University, Nagoya 464-8602, Japan}
\affiliation{Department of Physics, The University of Tokyo, Bunkyo-ku, Tokyo 113-0033, Japan}

\author{Kunimasa Miyazaki}
\affiliation{Department of Physics, Nagoya University, Nagoya 464-8602, Japan}

\author{Takeshi Kawasaki}
\affiliation{Department of Physics, Nagoya University, Nagoya 464-8602, Japan}

\date{\today}

\begin{abstract}
The reversible-irreversible (RI) transition of particle trajectories in athermal colloidal suspensions under cyclic shear deformation is an archetypal nonequilibrium phase transition which attracts much attention recently. Most studies of the RI transitions have focused on either dilute limit or very high densities well above the jamming transition point. The transition between the two limiting cases is largely unexplored. In this paper, we study the RI transition of athermal frictionless colloidal particles over a wide range of densities, with emphasis on the region below $\varphi_{\rm J}$, by using oscillatory sheared molecular dynamics simulation. We reveal that the nature of the RI transitions in the intermediate densities is very rich. As demonstrated by the previous work by Schreck {\it et al.} [Phys. Rev. E. {\bf 88}, 052205 (2013)], there exist the point-reversible and the loop-reversible phases depending on the density and the shear strain amplitude. We find that, between the two reversible phases, a quasi-irreversible phase where the particles' trajectories are highly non-affine and diffusive. The averaged number of contacts of particles is found to characterize the phase boundaries. We also find that the system undergoes the yielding transition below but in the vicinity of $\phi_{\rm J}$ when the strain with a small but finite strain rate is applied. This yielding transition line matches with the RI transition line separating the loop-reversible from the irreversible phases. Surprisingly, the nonlinear rheological response called ``softening'' has been observed even below $\varphi_{\rm J}$. These findings imply that geometrical properties encoded in the sheared configurations control the dynamical transitions.
\end{abstract}

\pacs{05.10.-a,61.43.-j,83.50.-v}


\maketitle
\section{Introduction} 

Athermal colloidal suspensions driven by cyclic shear are the simplest model system to study mechanical, rheological, and flow properties of the complex  
fluids. 
Recently, there have been many attempts to connect these macroscopic behaviors 
with the microscopic reversibility of trajectories of constituting particles.
If the amplitude of the oscillatory shear strain is small, the system finds an optimal configuration
after many cycles and the particles return to their original positions after every cycle (or several cycles), 
whereas beyond a critical amplitude, the trajectories become diffusive and irreversible.
This transition is referred to as the reversible-irreversible (RI) transition~\cite{Hinrichsen2000,Henkel2008book}.          

The RI transition has been studied mainly in two different
arenas: in the low density and high-density limit. In
the low-density limit, the transition from the reversible to
irreversible states is triggered by 
the instability of the trajectories due to the particle collisions and
hydrodynamic interactions when the strain amplitude exceeds a
critical value~\cite{Pine2005, Corte2008}.  
This transition is not the chaos transition but instead is a critical
phenomenon~\cite{Pine2005,Corte2008,Milz2013pre,Tjhung2015prl}. 
It is one of the nonequilibrium phase transitions
in which the order parameters, relaxation time, and the correlation
lengths obey power laws whose exponents are very close to those of the Directed
Percolation (DP) universality class of the absorbing state transitions. 
Now the RI transition becomes an epitome of the nonequilibrium phase transition, along with other
diverse examples, such as contact
processes~\cite{Hinrichsen2000,Henkel2008book}, a topological transition
of the liquid crystal turbulence~\cite{Takeuchi2007}, the sheared-vortex
in a superconductor~\cite{Okuma2011prb}, the skyrmions in chiral magnets~\cite{Brown2019njp}, and the laminar-to-turbulent
transition in simple fluids~\cite{Sano2016natp}.   

If the particles are soft, the RI transition also takes place at the very high density region well
above the jamming transition density $\phi_{\rm J}$, where most particles are in
contact with each
other~\cite{Lundberg2008pre,Schreck2013pre,Regev2013pre,Fiocco2013pre,Keim2013sm,Keim2014prl,Nagamanasa2014pre,Regev2015natc,Priezjev2013pre}.   
Even though the particles experience multiple collisions during
one oscillatory cycle and the trajectories are complicated, 
there exists a reversible phase before the system enters to the irreversible phase at
larger strain amplitudes~\cite{Lundberg2008pre,Schreck2013pre,Regev2015natc}. 
Contrary to the low-density limit, this RI transition is discontinuous~\cite{Knowlton2014sm,Kawasaki2016pre,Tjhung2017pre}; 
across the critical strain amplitude, the order parameter such as the number of the
irreversible particles becomes finite discontinuously.  
Furthermore, the critical amplitude of the RI transition is almost identical to
that of the yielding transition, another nonequilibrium phase transition
from the elastic amorphous phase to the plastic and flowing phase. 

Although the two limiting cases of low- and high-densities are relatively well investigated, 
attempts to bridge seamlessly the two regimes are relatively few. 
Most challenging is the vicinity of the jamming transition point, $\varphi_{\rm J}$, where the three
distinct transitions of the jamming, yielding, and the RI
transitions meet~\cite{Lavrentovich2017pre,Dagois-Bohy2017sm}.
Even less explored is the intermediate density region between the low-density limit and $\varphi_{\rm J}$. 
Recently, Schreck {\it et al.} have studied the RI transition for a broad range of densities and
strains below $\phiJ$~\cite{Schreck2013pre}. 
They found three distinct phases, {\it i.e.}, 
the irreversible phase slightly below $\phiJ$, 
the point-reversible phase at low strain amplitudes, where the particles do not collide during the
oscillatory cycles, 
and the loop-reversible state between the point-reversible and
irreversible phases, where the trajectories are highly non-affine but still most of the particles
reversibly come back to the original position after every (or several) cycle(s). 
On the other hand, the relationship between the rheological properties
and the particle configurations slightly below $\phiJ$ is investigated by Vinutha {\it et
al.}~\cite{Vinutha2016natp,Vinutha2016jsmte}, 
who  found that the stationary configurations of the frictionless particles under
quasistatic uniform shear are analogous to those of the shear-jammed frictional
particles. 
These results suggest that the nature of the RI transition at the intermediate density
region is very rich and can not be described as a mere extrapolation
from the low-density region. 
Related results in three dimensions are presented in \cite{Das2019prep}.

In this paper, we perform simulations for a two dimensional harmonic potential
system under oscillatory shear strain for a wide range of densities near
the jamming transition point.
We find varieties of nonequilibrium phase transitions.
At the low-density side $\phi < \phi_{\rm J}$, the RI transition falls into neither the
DP-like continuous transition nor the discontinuous transition.  We investigate the close
relationship of the RI transition with both the mechanical and rheological properties on the
one hand and the geometrical proprieties of the particle configurations such as the contact
networks on the other hand.  
We examine the oscillatory strain with both zero (or quasi-static limit)
and finite-frequencies. 
It is found that the onset of the RI transition 
is insensitive to the frequencies, 
whereas mechanical properties appreciably change with frequencies.
The RI transition at the high-density side $\phi > \phi_{\rm J}$ is discontinuous but 
the one-cycle displacement of the particles which plays a role as the order parameter vanishes as
the system approaches $\phi_{\rm J}$ from above.
As the density decreases further, the discontinuous RI transition line continues 
even below $\phi_{\rm J}$ while the yielding transition line terminates at $\phi_{\rm J}$.
Interestingly, when the small but finite frequency cyclic shear is applied, 
the yielding transition is observed even below $\varphi_{\rm J}$ and 
exactly at the RI transition point.
We deliberately avoid studying the system exactly at $\phi_{\rm J}$, 
where the jamming criticality intervenes and analysis becomes extremely challenging.

The paper is organized as follows. 
We explain the simulation methods in Section \ref{sec:method}. 
In Section \ref{sec:results}, the order parameter of the RI transition for several densities is
analyzed and we construct the nonequilibrium phase diagram. 
We also discuss  the relationship of the phase diagram with the geometrical properties of particle 
trajectories and rheology of the system. 
Finally, we summarize and give some remarks in Section \ref{sec:conclusions}.

\section{Simulation method} 
\label{sec:method}

The system we study is a two-dimensional equimolar binary mixture of frictionless particles 
with diameters $\sigma_{\rm L}$ and $\sigma_{\rm S}$, interacting with a harmonic potential. 
The size ratio of small and large particles is $\sigma_{\rm L}/\sigma_{\rm S}= 1.4$. 
We investigate the systems for a wide range of densities $0.70 \leq \varphi \leq 1.0$. 
The jamming transition density in our system is $\varphi_{\rm J} \approx 0.842$~\cite{Lerner2012pnas}. 
We use the two protocols to simulate the system. 
The first is the finite-frequency oscillatory shear protocol, where the particles are driven by the
overdamped equation with the Stokes' drag~\cite{Durian1995,allen1987}.  
The equation of motion is given by
\be 
\zeta_{\rm s} \left[ \odif{{\bf r}_i}{t} - 
\dot{\gamma}(t)y_i(t){\bf e}_x \right] +
\sum_{j} \frac{\partial U( r_{ij} )}{\partial 
{\bf r}_{i}} = 0,
\label{em}
\ee
where $\zeta_{\rm s}$ is a friction constant, 
$r_{ij}= |{\bf r}_i - {\bf r}_j|$ 
is the interparticle distance between the $i$-th and $j$-th particles, 
${\bf e}_x=(1,0)$ is the direction of the shear, 
and $\dot\gamma(t)$ is the shear rate. 
The interaction potential is given by $U(r_{ij})=\frac{\epsilon}{2}
\left(1-{r_{ij}}/{\sigma_{ij}}\right)^2\Theta (\sigma_{ij}-r_{ij})$, where $\epsilon$ is an energy scale,
$\sigma_{ij}=(\sigma_i+\sigma_j)/2$ is the interparticle distance at
contact, and $\Theta (x)$ is the Heaviside step function.  
In our simulations, we use $\sigma_{\rm S}$, $\tau_0 = \sigma_{\rm S}^2
\zeta_{\rm s} /\epsilon$, $\epsilon$, and 
$\epsilon/\sigma_{\rm S}^3$ as the unit of the length, time, energy, and the stress, respectively. 
We apply an oscillatory deformation using the Lees-Edwards periodic
boundary condition~\cite{allen1987}. 
The time evolution of the strain is given by ${\gamma}(t)= \gamma_0
[1-\cos(\omega t)]$, where $\gamma_0$ is the amplitude of the imposed shear strain and $\omega = 2 \pi/T$ is the frequency of the oscillation. 
We compute an initial configuration as a random distribution at $t=0$.
The oscillation period $T$ is chosen to be very large and we use mainly $T =10^4 \tau_0$. 
We solve Eq.~(\ref{em}) discretized with the semi-implicit Euler's algorithm. 
The time-step is chosen to be $\Delta t=0.1$.  
We have checked that using the higher order discretization algorithm such as the Heun method does not
alter the results~\cite{allen1987}, because our simulations are driven slowly enough with the
overdamped equations of motion and the accumulation of the rounding-off error is suppressed.   
We also note that the velocity Verlet method, which is used in a micro-canonical molecular dynamics
simulations, is not applicable in our simulations, because our system is fully overdamped.  
The system size is $L=20$ for the results presented in the main text. 
We confirm that the main results do not qualitatively change with the
larger system size of $L=40$. 
The maximal simulation time is set to $t_{\rm sim}=4000T$. 
To improve the statistics, we perform at least five independent runs in each of the parameter sets
of $(\phi, \gamma_0)$. 
We use the bracketing symbol $\langle \cdots \rangle $ for ensemble
averages of variables discussed below.   

The second is the athermal quasi-static (AQS) protocol which corresponds
to the small strain rate limit $\dot{\gamma}\rightarrow 0$. 
We first give the small shear strain
at each step to drive the particles toward the shear direction by  
\be
{\bf r}'_j(n+1) = {\bf r}_j(n) + \Delta \gamma(n) y_j(n) {\bf e}_x,
\ee
where ${\bf r}_j(n)$ is the position of the $j$-th particle at the $n$-th simulation step. 
After one step, the position of the particles,  ${\bf r}'_j(n+1)$, are
relaxed by the conjugate gradient line-search algorithm to minimize the energy. 
The shear strain evolves with $\gamma(n+1) = \gamma(n)+ \Delta\gamma(n)$, where 
$\Delta \gamma(n) = 2\gamma_0\pi/N_{\rm cycle}\sin{(2\pi n/N_{\rm cycle})}$.  
In this study, we set a period of the simulation steps per cycle to 
$N_{\rm cycle}=10^4$ which is sufficiently large.

\section{Results}
\label{sec:results}

\subsection{Dynamics and Phase Diagram}

\begin{figure}
\includegraphics[width=1.0\columnwidth,angle=-0]{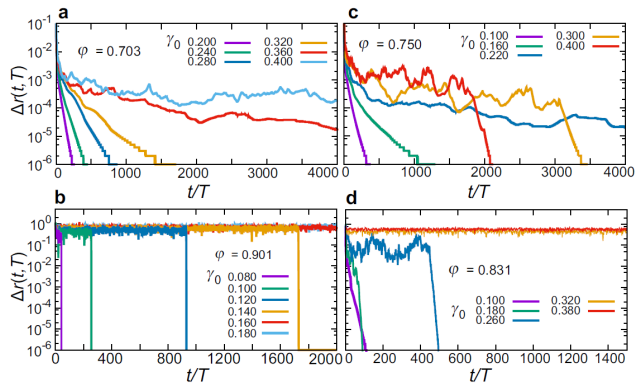}
\caption{\label{fig1} 
 The time evolution of the averaged one-cycle displacement $\Delta r(t,T)$.
(a): At $\phi = 0.703$ far below $\varphi_{\rm J}$.
For $\gamma_0 \lesssim 0.40$, $\Delta r(t,T)$ {\it continuously} drop to zero after a timescale $\tau_{\rm L}$ (see the main text for the definition of $\tau_{\rm L}$).
(b): At $\varphi = 0.901$ far above $\varphi_{\rm J}$.  
For $\gamma_0 \lesssim 0.14$, $\Delta r(t,T)$ {\it discontinuously} drop to zero. 
(c): At $\varphi = 0.750$. 
For small $\gamma_0$ ($\gtrsim 0.2$), $\Delta r(t,T)$ relaxes to zero exponentially with a power law
 tail, whereas, for $\gamma_0 \sim 0.2$, $\Delta r(t,T)$ first relaxes to a finite value before it
 discontinuously drop to zero. 
(d): At $\varphi = 0.831$ slightly below $\varphi_{\rm J}$.
For $\gamma_0 \lesssim 0.30$, the displacements {\it semi-discontinuously} drop to zero.}
\end{figure}
\begin{figure}[t]
\includegraphics[width=0.9\columnwidth,angle=-0]{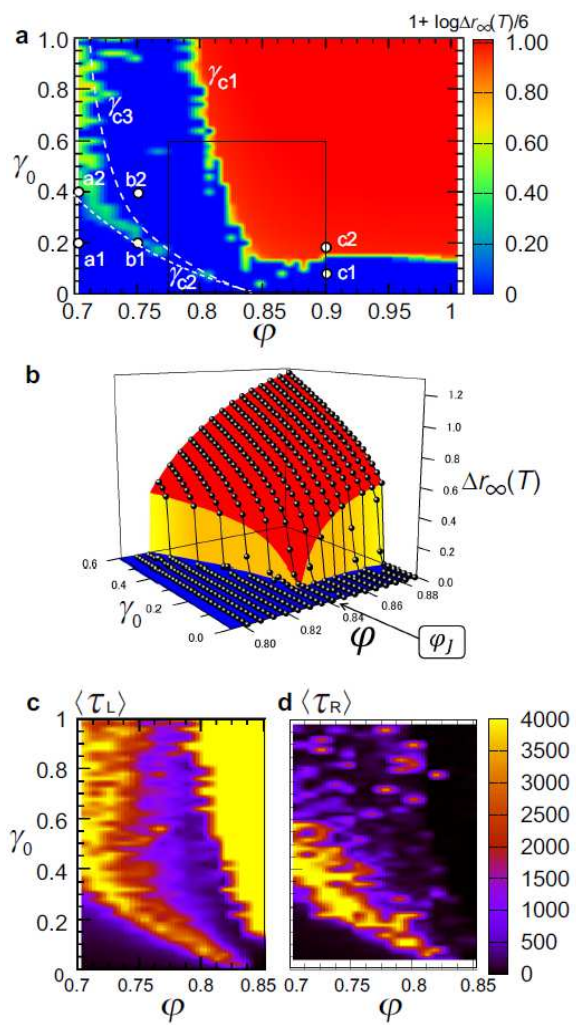}
\caption{\label{fig2} 
 The phase diagrams of the RI transition shown as the color map of several observables.
(a): The phase diagram obtained from $\Delta r_{\infty}(T)$ in the steady state. 
The color field presents the value of $1+\log{\Delta r_{\infty}(T)}/6$ (see the color bar).
$\gamma_{\rm c1}$ (discontinuous),  $\gamma_{\rm c2}$ (continuous), and 
 $\gamma_{\rm c3}$ (discontinuous) indicate the locations of the phase boundaries (see the main text for detail). 
The two dashed lines, $\gamma_{\rm c2}$ (lower bound) and
$\gamma_{\rm c3}$ (upper bound),  are guides to the eye.
(b): Bird's-eye view of $\Delta r_{\infty}(T)$ as a function of $\phi$ and $\gamma_0$ for the
 region marked by a black square in (a). The colored areas are the guide to the eye.
(c): The phase diagram drawn using the averaged lifetime of irreversible trajectories
 $\langle\tau_{\rm L}\rangle$.
(d): The averaged relaxation time $\langle\tau_{\rm R}\rangle$. 
For (c) and (d), the region above $\phi > 0.85$ is not shown because it is basically identical with
 those of (a). 
}
\end{figure}

First, we introduce the averaged particle displacements for one cycle defined by
\be
\Delta r (t,T) = \frac{1}{N}\sum_{j=1}^N |{\bf r}_j(t+T)-{\bf r}_j(t)|.
\label{disp}
\ee
This characterizes the reversibility of the particle trajectories and is often used as an order
parameter of the RI transition. 
$\Delta r (t,T) \sim 0$ if the system is in the reversible state 
and it is finite in the irreversible state.  
We regard the system is in the reversible state when $\Delta r (t, T) < 10^{-6}$. 
We measure $\Delta r(t,T)$ for
various values of both $\varphi$ and $\gamma_0$. 
To cover the broad range of the parameter space of $(\varphi, \gamma_0$) efficiently, we
use the finite frequency protocol with a large $T$. This protocol is much faster than that of the
AQS.  
As we discuss below, the frequency dependence of $\Delta r(t,T)$, and thus the phase diagram, 
 is weak as far as $T$ is large enough. 
We switch to the AQS protocol when we compute the geometrical and mechanical properties
because they sensitively depend on the frequency. 
In Figure~\ref{fig1}, we show typical time evolutions of $\Delta r(t,T)$ for four representative
densities below and above $\phiJ$ and for several $\gamma_0$'s.  
Figure~\ref{fig1}{\bf a} is the results at a low density, $\varphi = 0.703$, far below $\phi_{\rm J}$.  
For $\gamma_0 \lesssim 0.40$, the displacements exponentially relax to
zero.  As $\gamma_0$ increases, the relaxation slows down, developing power-law tails, and
eventually the system enters the irreversible state with finite $\Delta r(t,T)$'s.
The result for a high density, $\varphi = 0.901$, far above $\varphi_{\rm J}$ is shown in
Figure~\ref{fig1}{\bf b}.
Relaxation behavior is qualitatively different from that at low densities.  
$\Delta r(t,T)$ starts from the irreversible (active) state with a finite value.
It remains almost constant until it {\it discontinuously} drops to zero or enters to the
reversible (absorbing) state. 
The heights of the plateau of $\Delta r(t,T)$ are insensitive to $\gamma_0$ 
but the relaxation time in the irreversible state increases sharply as $\gamma_0$ increases 
before it eventually exceeds the time window of our simulation 
at $\gamma_{\rm c} \approx 0.14$. 
The previous study has shown that $\gamma_{\rm c}$ where the relaxation time diverges 
matches with the yielding transition point $\gamma_{\rm Y}$ at which 
the system loses the elasticity and starts flowing~\cite{Kawasaki2016pre}. 
We observed,  at high densities,  that a fraction of trajectories are 
multi-period reversible. In other words, some particles come back to their original
positions only after several cycles. 
In the present study, we do not include the data for the analysis.

The results for intermediate densities between the low and high densities are surprisingly rich. 
Figure~\ref{fig1}{\bf c} shows the results at $\varphi = 0.750$. 
If $\gamma_0$ is small, the relaxation behavior of $\Delta r(t,T)$ is qualitatively the same as those
at $\varphi=0.703$ and it decays exponentially. 
As $\gamma_0$ increases, $\Delta r(t,T)$ starts developing the power law tails. 
The relaxation time increases and eventually becomes larger than the simulation windows.  
As $\gamma_0$ increases further, however, this trend is reversed and 
the relaxation time becomes shorter. 
In other words, the system shows the reentrant behavior.
After this reentrance, the relaxation behavior qualitatively changes and becomes 
more like that observed above $\varphi_{\rm J}$ (Fig.~\ref{fig1}{\bf b});  
it first quickly decays to a plateau, stay there for  a long time,  and then abruptly and discontinuously 
drop to zero.
Contrary to the results at $\phi> \phiJ$, the heights of the plateau increase continuously 
and the relaxation time shortens as $\gamma_0$ increases.  
This reentrant transition of the relaxation time below $\varphi_{\rm J}$ is already 
reported by Schreck {\it et al.}~\cite{Schreck2013pre}. 
They found that this transition separates the two qualitatively
different reversible phase which is
referred to as the point-reversible and loop-reversible phases. 
We shall analyze thoroughly the dynamical and geometrical properties of these phases in the next
subsection. 

Figure~\ref{fig1}{\bf d} is the result at $\varphi = 0.831$, slightly below $\phiJ$. 
In this region, we again observe the exponential relaxation of $\Delta r(r, T)$ at very small 
$\gamma_0$ ($=0.1$). 
As $\gamma_0$ increased, however, the relaxation time barely increase but instead it develops the
plateau with the finite heights and its behavior becomes more like that observed
above $\phiJ$ (see Fig.~\ref{fig1}{\bf a}), {\it i.e.}, a quick decay to a plateau and sudden and step-wise
relaxation to zero. 
For $\gamma_0  \gtrsim 0.3$, $\Delta r(t,T)$ remains constant in our simulation
windows, suggesting that the system entered the irreversible state. 
It is interesting to observe a qualitatively similar behavior to that above $\varphi_{\rm J}$  even
in the unjammed phase. 
In the next subsection, we show that the RI transition observed in this density regime is closely
related to the macroscopic rheology.

Using these simulation results, we draw the {\it dynamical phase diagram} of the RI
transition using $\Delta r(t, T)$ in the stationary state 
as the order parameter for a broad range of $\phi$ and $\gamma_0$. 
We run the long simulations up to $t=4000 T$ and judge that the system is in the stationary state 
if $\Delta r(t,T)$ becomes independent of time.  
We denote the ensemble average of $\Delta r(t=4000T,T)$ as $\Delta r_{\infty} (T)$. 
Note that the diffusion constant $D$ is an alternative candidate
of the order parameter. In Ref.~\cite{Kawasaki2016pre}, it is shown that the position of the RI transition is not affected by choice of the order parameters if the period-doubling data are excluded.
Figure~\ref{fig2}{\bf a} is the phase diagram drawn as the contour plot of 
$\Delta r_{\infty} (T)$ as a function of $(\varphi, \gamma_0)$. 
In order to enhance the visibility of the small values of $\Delta r_{\infty} (T)$, 
$1 + \log \Delta r_{\infty} (T)/6$ is used to define the color map. 
The reversible region where $\Delta r_{\infty} (T)=0$ (or more precisely $\Delta r_{\infty} (T)< 10^{-6}$) 
is colored by blue and the largest value of $\Delta r_{\infty} (T)=1$ is colored by red (see the color
column).   
In order to draw this figure, we simulated the system for the range of 
$0.7 \leq \phi \leq 1.0$ and $0 \leq \gamma_0 \leq 1.0$ for grid points
separated by $\delta \varphi=0.05$ and $\delta \gamma_0=0.05$ 
for most cases. 
The finer grid sizes of $\delta \varphi=0.01$ and $\delta \gamma_0=0.02$
are used in the vicinity of the phase boundaries. 
This phase diagram reflects the dynamic properties of $\Delta r(t,T)$ observed in Fig.~\ref{fig1}. 
There are several qualitative features which draw our attention. 

Firstly, in the high-density side above $\phi \gtrsim 0.8$, one observes
an irreversible phase colored by red which is sharply separated from the
reversible phase in blue.
$\gamma_{\rm c1}$ represents the phase boundary line.
Above $\varphi_{\rm J}\approx 0.842$, the irreversible phase is always observed
and the transition point $\gamma_{\rm c1}\approx 0.15$ is almost constant over all densities above 
$\phiJ$. 
The abrupt change of the colors at $\gamma_{\rm c1}$ is a consequence of a sudden and discontinuous 
increase of $\Delta r_{\infty} (T)$~\cite{Kawasaki2016pre}. 
The value of $\Delta r_{\infty} (T)$ in the irreversible phase is very close to 1 irrespective of 
$\varphi$ and $\gamma$. 
$\gamma_{\rm c1}$ is not exactly constant. 
As $\phi$ is decreased from above, it is slightly bent upward around 
$\varphi \approx 0.9$ and then downward as $\phiJ$ is approached, 
before it turns upward sharply at the edge at $\phi=\phiJ$. 
$\gamma_{\rm c1}$ increases sharply at $\varphi < \phiJ$. 
The presence of the discontinuous transition below $\phiJ$ is also
observed in Ref.~\cite{Schreck2013pre}. 
In the large $\gamma_0$ limit, this RI transition line seems to diverge at $\varphi \approx 0.8$. 
Note that this value is very close to the random loose packing density where 
the particles would undergo the mechanical transition if there is an
interparticle frictional force. This result is reminiscent of 
the finding that configurations of the sheared frictionless spheres 
are similar to those of the shear jammed system of the frictional spheres reported in
Ref.~\cite{Vinutha2016natp}.  
$\Delta r_{\infty} (T)$ at $\phi < \phiJ$, however,  is quantitatively
different.  
Its magnitude is not an order of unity but decreases 
as the density is decreased towards the phase boundary for a fixed
$\gamma_0$. 
Figure~\ref{fig2}{\bf b} is the birds eye's view of 
$\Delta r_{\infty}(T)$ near $\phiJ$. 
It shows that the RI transitions of both sides of $\phiJ$ are discontinuous but 
$\Delta r_{\infty} (T)$  tends to vanish at the $\phi=\phiJ$, while  
$\gamma_{\rm c1}$ there remains unchanged. 
It is premature to conclude that this vanishing order parameter is the sign
of the continuous transition. It is because the resolution of the
parameters is too low and
assessing the critical behavior of $\Delta r_{\infty}(T)$
at  $\phiJ$ exactly  is challenging due to a subtle interplay of the 
criticalities of the jamming and RI transitions~\cite{Lavrentovich2017pre,Dagois-Bohy2017sm}.   
We shall revisit this issue in future work.

\begin{figure*}[t]
\includegraphics[width=0.9\textwidth,angle=-0]{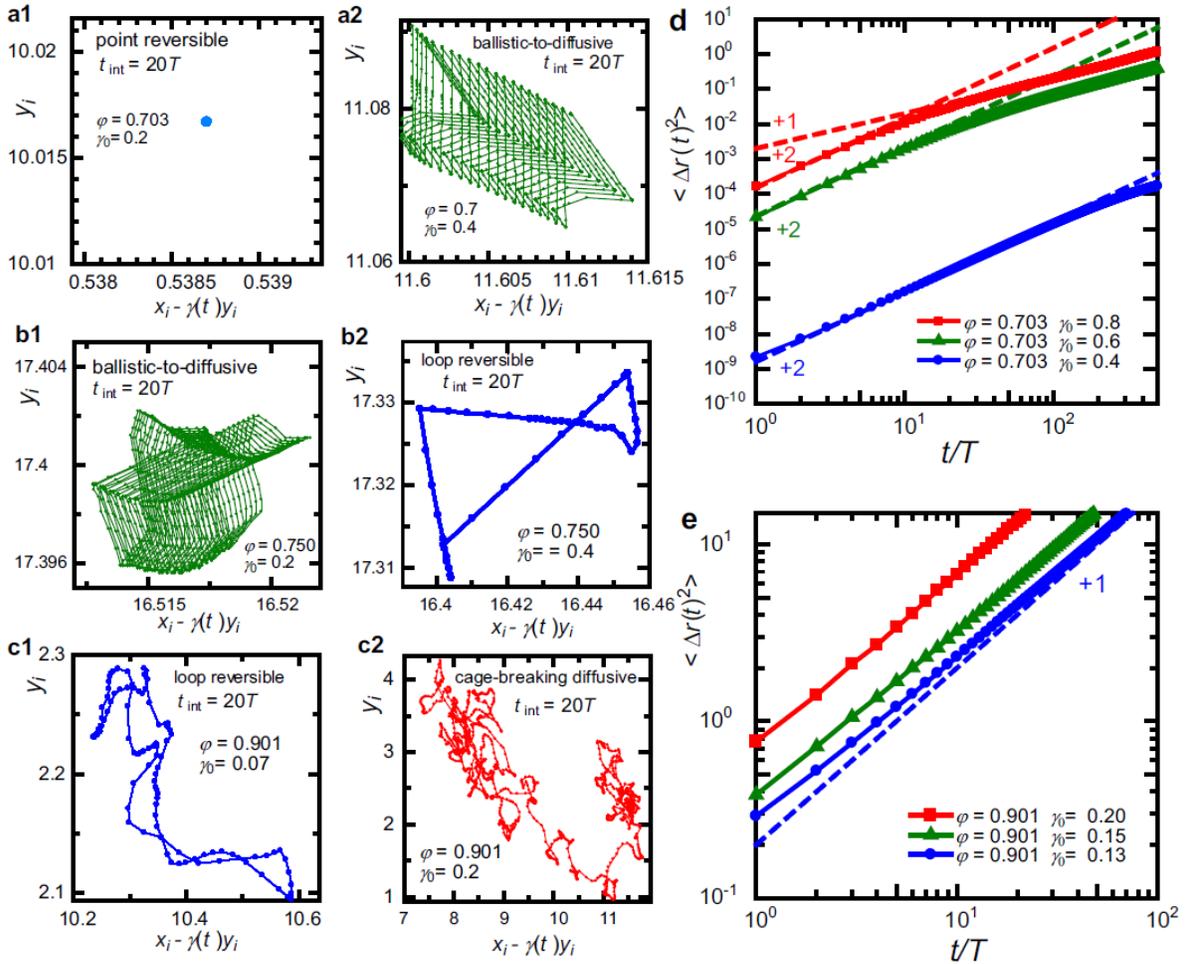}
\caption{\label{fig3} 
(a1)--(c2): Representative particle trajectories during 20 cycles.
 The symbols a1--c2 corresponds to those marked in Fig.~\ref{fig1}{\bf a}.
Depending on $\gamma_0$ and $\phi$, the topology of trajectories qualitatively change. 
(a1): A point reversible trajectories.
(a2) and (b1): Ballistic-diffusive trajectories.
(b2) and (c1): Loop-reversible trajectories.
(c2): Cage-breaking diffusion.
The mean squared displacements (d) at a small density, $\varphi = 0.703$,  
 with $\gamma_{0} = 0.4$, 0.6, 0.8 
and (e) at a high density $\varphi = 0.901$ with $\gamma_{0} = 0.13$, 0.15, 0.20. 
}
\end{figure*}

Secondly, and most noticeably, there is a ``peninsula'' of the
irreversible phase below $\phiJ$, where 
$\Delta r_{\infty} (T)$  is finite in the low-density and
small-$\gamma_0$  region, surrounded by the reversible states. 
The presence of this irreversible region is a manifestation of the reentrant behavior of $\Delta r(t, T)$ shown in
Fig.~\ref{fig1}{\bf b}.
As we shall discuss more in detail in the next subsection, this region
is surrounded by the two phase boundaries 
on the low $\gamma_0$-side (denoted as $\gamma_{\rm c2}$) and upper $\gamma_0$-side (denoted as
$\gamma_{\rm c3}$).
We find that the RI transition across $\gamma_{\rm c2}$ is continuous whereas it is discontinuous
at $\gamma_{\rm c3}$. 
As the density increases, the two phase boundaries of $\gamma_{\rm c2}$ and $\gamma_{\rm c3}$ tend to converge
to zero around $\phiJ$. 
The phase boundaries of the irreversible phase, 
especially $\gamma_{\rm c3}$, is jerky and blurred 
due to very large sample-to-sample fluctuations of $\Delta r_{\infty}(T)$. 
One may be tempted to consider this irreversible peninsula is a metastable state 
which vanishes in the large time limit ($t\rightarrow \infty$) or in the
AQS limit ($T \rightarrow \infty$).
However, we verify that this phase is unexpectedly stable and
the dynamical behaviors of the particle trajectories in this phase 
are 
distinct from that outside of the irreversible phase.  
We also verify that this phase survives in the AQS limit. 

Instead of the order parameter $\Delta r_{\infty} (T)$, 
one can employ the relevant time scales as the order parameters 
to draw the phase diagram. 
The phase boundaries are drawn as the points 
at which the time scales diverge. 
Note that there are two types of time scales in our systems depending on the
nature of the RI transitions.
If the transition is continuous as we observe at low densities, 
$\Delta r_{\infty} (T)$ is well described by an exponential function 
from which one can define the relaxation time $\tau_{R}$.
To define $\tau_{\rm R}$, we use an empirical function given by ~\cite{Corte2008}
\be
\Delta r (t,T) = \Delta r_{0}\frac{e^{-t / \tau_{\rm R}}}{t^{\delta}}, 
\label{fit_curve1}
\ee 
where $\Delta r_0 \equiv \Delta r(t=0, T)$. 
On the other hand, if the transition is discontinuous as we observe at $\gamma_{\rm c1}$ and
$\gamma_{\rm c3}$ in Fig.~\ref{fig2}{\bf a},  
$\Delta r_{\infty} (T)$ develops a plateau. 
There are initial relaxations towards the plateau and the final abrupt drops from the plateau
to the reversible phase. 
We refer the time scale of the latter process as the life time $\tau_{\rm L}$ in order to
discriminate the relaxation time $\tau_{\rm R}$ of the initial relaxation. 
$\tau_{\rm R}$ is obtained by fitting by the generalized expression of
Eq.~(\ref{fit_curve1}) given by 
\be
\Delta r (t,T) = (\Delta r_{0} - \Delta r_{\rm s})\frac{e^{-t / \tau_{\rm R}}}{t^{\delta}} + \Delta r_{\rm s},
\label{fit_curve2}
\ee 
where $\Delta r_{\rm s}$ is the plateau value. 
On the other hand, we define the life time $\tau_{\rm L}$ as the time at which 
$\Delta r(\tau_{\rm L},T)$ drops to $10^{-6}$.
Figure~\ref{fig2}{\bf c} and {\bf d} are the phase diagram or the contour plots of 
the averaged values of $\tau_{\rm L}$ and $\tau_{\rm R}$.
The region above $\phiJ$ is not shown because it is identical to that for $\Delta r_\infty(T)$. 
In the vicinity of the phase boundary $\gamma_{\rm c2}$, 
the iso-$\tau_{\rm L}$ line coincides with the iso-$\tau_{\rm R}$ line,
since $\Delta r_{\infty}$ continuously decreases to 0 toward the critical point.
On the other hand, near $\gamma_{\rm c3}$, the two time scales are decoupled. 
While the region where $\tau_{\rm L} > 4000T$ (yellow colored region in Fig.~\ref{fig2}{\bf c}) 
matches with the irreversible peninsula in Fig.~\ref{fig2}{\bf a},  
The large $\tau_{\rm R}$ ridge only exists along $\gamma_{\rm c2}$-line (see Fig.~\ref{fig2}{\bf d}). 
This result bolsters that the phase boundaries of the peninsula of the irreversible phase 
is delineated by the continuous transition at small $\gamma_{0}$'s 
and discontinuous transition at large $\gamma_0$'s. 
Schreck {\it et al.}~\cite{Schreck2013pre} have demonstrated that 
this diverging time scale separates the reversible phase into the two distinct phases; 
the point-reversible phase where the one-cycle trajectory is completely affine-like and 
the loop-reversible phase where it is not. 
Our results suggest that there exist a (meta-)stable irreversible phase between the point- and
loop-reversible phase. 
In the next subsection, we characterize this distinct phase using the topology of the one-cycle
trajectories and their long time dynamics.

\subsection{Particle trajectories}

The presence of the region of the large relaxation time in
Fig.~\ref{fig1}{\bf a} is hardly the evidence that this peninsula is 
the {\it bona fide} irreversible phase. 
To identify the nature of this exotic state, 
we characterize the microscopic trajectories of particles inside and the
vicinity of the peninsula in more detail. 
We analyze the microscopic trajectories of particles in several representative state points in the
phase diagrams. 
Figures~\ref{fig3}{\bf a1}-{\bf c2} are the trajectories of a single particle for
multiple cycles for the state points marked in Figure \ref{fig1}{\bf a}. 
We plot the non-affine trajectories, $\rNA_j(t)\equiv (x_j(t)-\gamma(t)y_j (t), y_j(t))$, where the affine part of the displacement 
subtracted from the coordinate. 
Here, we used the finite frequency protocol with $T=10^4$.  
We have also carried out the same analysis using the AQS protocol ($T=\infty$) and found that the
results do not qualitatively change.
Dots which are connected by lines in each figure represent the position of the particle at each
incremental step of strain for $T/100$. 

If both $\varphi$ and $\gamma_0$ are very small, the trajectory is just a single point 
as shown in Fig.~\ref{fig3}{\bf a1} because the particle is convected by the shear 
and, after a cycle, comes back to
the original position without a collision with other particles. 
This reversible state is referred to as the point-reversible state~\cite{Schreck2013pre}. 
This single point state is observed over the region where $\phi < \phiJ$ and $\gamma <
\gamma_{\rm c2}$. 

For $\phi> \phiJ$,  on the other hand, 
the trajectories of the reversible phase are qualitatively different 
from the point-reversible trajectories, as shown in Figure~\ref{fig3} {\bf c1}. 
Each particle travels complicated non-affine trajectories until it
comes back to the original position after one cycle.  
This is called the loop-reversible state.
If $\gamma_0$ is increased above $\gamma_{\rm c1}$ at a fixed $\phi$($> \phiJ$), 
the trajectory fails to close the loop after a cycle and the system becomes irreversible 
as demonstrated by previous studies~\cite{Kawasaki2016pre}. 
We call this state the loop-irreversible (Fig.~\ref{fig3}{\bf c2}).  
This RI transition point coincides with the yielding transition
($\gamma_{\rm c}=\gamma_{\rm Y}$)~\cite{Knowlton2014sm,Kawasaki2016pre}.  
As $\phi$ is approached to $\phi_{\rm J}$ from above,  
the trajectories become more complicated and some particles come back to
the original positions only after several cycles~\cite{Lavrentovich2017pre}. 
The particle trajectories below and above $\gamma_{c1}$ below $\phiJ$,
where the yielding transition is absent, are qualitatively the same as
those above $\phiJ$.

Striking are the behaviors of the trajectories at $\phi < \phiJ$ and $\gamma_0 > \gamma_{\rm c2}$. 
In the reversible phase above the peninsula, {\it i.e.}, $\gamma_0 >
\gamma_{\rm c3}$, the trajectories are again the loop-shaped but their shape is somewhat simpler than those observed at $\phi > \phiJ$.
Figure~\ref{fig3}{\bf b2} is the typical loop trajectory observed at $\phi=0.75$ and 
$\gamma_{0}=0.4$ which lies in the reversible phase well above $\gamma_{\rm c3}$. 
We observed a variety of loop trajectories in this reversible phase, 
whose shapes vary depending on the locations in the parameter space.
Inside the peninsula of the irreversible phase surrounded between $\gamma_{\rm c2}$ and 
$\gamma_{\rm c3}$, the trajectories are very different. 
Figures~\ref{fig3}{\bf a2} and {\bf b1}) are the representative trajectories which are observed 
in this irreversible phase.
The particles' trajectories in these figures are recorded over 20 cycles. 
The particles trace loop-like trajectories in each cycle but, at the very end of one cycle, 
they always fail to close the loop. 
After every cycle, the endpoints migrate by a very small distance and 
they depart from the original position. 
After many cycles, the trajectories eventually becomes diffusive. 
This crossover of the particle dynamics from a quasi-reversible to diffusive 
is most clearly demonstrated by evaluating the mean squared displacements (MSDs) defined by 
\be
\ave{\Delta r^2(t)} = \frac{1}{N} \ave{\sum_{j=1}^N 
\left\{\rNA_j(t_0+t)-\rNA_j(t_0)\right\}^2},
\label{msd}
\ee
where $N$ is the total number of particles, $\rNA_j(t)$ is the non-affine displacement of the
$j$-th particles. 
$t_0$ is chosen to be longer than $\tau_{\rm R}$, so as to ensure the
system is in the stationary state. 
Figures~\ref{fig3}{\bf d} and \ref{fig3}{\bf e} are the MSDs below and above
$\phiJ$ for several $\gamma_0$, respectively.
Note that $\ave{\Delta r^2(t)}=0$ if the system is in the reversible state,
as we measure the trajectories in the stationary state. 
The MSD for $\varphi > \varphi_{\rm J}$ (Fig.re~\ref{fig3}{\bf e})
is subdiffusive at short times, $\ave{\Delta r^2 (t)} \propto t^{\alpha}$ with
$\alpha <1$, but after several cycles they become diffusive, $\ave{\Delta r^2 (t)} \propto t$.  
This subdiffusive behavior is the reflection of the particles motion hindered by the 
multiple collisions with surrounding particles like in the supercooled liquids near the glass
transition point. 
We call this motion {\it the cage-breaking} diffusion. 
On the other hand, 
Figure~\ref{fig3}{\bf d} is the MSD in the irreversible phase at 
$\varphi < \varphi_{\rm J}$ (in the peninsula).
Although the MSD becomes diffusive eventually, the short time behavior
is ballistic first,  {\it i.e.}, $\ave{\Delta r^2 (t)} \propto t^2$ for
a long time. 
We refer to this behavior as {\it ballistic diffusive}.
This ballistic motion is a direct reflection of the small but systematic migration of the endpoints
shown in the trajectory in Fig.~\ref{fig3}{\bf a2} and \ref{fig3}{\bf b1}. 
The crossover time from the ballistic to diffusive regime agrees well with $\tau_{\rm R}$ and,
as $\gamma_0$ is lowered to $\gamma_{\rm c2}$ from above, it becomes longer and 
eventually goes out from the simulation window. 
Again we emphasize that the qualitative behaviors of the trajectories
are insensitive to the frequencies of the oscillation and is observed in the AQS protocol as well. 
Finally, to reassure that the irreversible peninsula state is robust 
insensitive to the protocol to prepare the system and 
the initial configuration of the particles,
 we prepared a configuration obtained in the loop-reversible states at large $\gamma_0$'s and then 
carried out the same simulations.  
We found that the shape of the trajectories remain qualitatively unchanged and
$\Delta r(t,T)$ are identical to those obtained from the random configurations. 
These results bolster that the irreversible state which we observed is the genuine stable
irreversible phases characterized by the continuous transition from the point-reversible to
irreversible phase and the discontinuous transition from the irreversible to loop-reversible
phase, as $\gamma_0$ is increased for a fixed $\phi$.
Such an exotic reentrant transition is a consequence of the emergence of the complex geometrical
landscape. 
At the intermediate $\phi$'s and large $\gamma_0$'s, 
there is a dynamical phase where the particles can find the way back to the original position 
after long sojourns in the valleys of the complex geometrical landscape
and multiple collisions with surrounding particles.
Many cycles of slow or quasi-static shear cycles can train the system
and encode and memorize the
information of the optimized pathways for the particles to stay in their stable configurations. 
However, if $\gamma_0$ is not large enough, 
particles cannot trace the encoded pathways and complete the loop 
in one cycle and thus fail to come back to their original positions. 
In the next subsection, we consider the relationship of the properties of the phase diagram
with the geometrical properties of the sheared configurations of the particles. 

\subsection{Geometry of the Particle configurations}
Since the system we study is athermal, any nonequilibrium transition of the trajectories must
be related to the geometrical properties imprinted in the configuration of the particles.  
To characterize them, we evaluate the contact number $Z$ of adjacent particles.
In principle, $Z$ can be obtained by integrating the radial distribution function $g(r)$ over the distance of the first coordination
shell.  
Contrary to the overall shape of the phase diagram and qualitative behaviors of the trajectories,
the contact number is very sensitive to the frequency of the oscillatory shear. 
Therefore, we employ the AQS protocol to draw the iso-$Z$ lines. 
We use the method proposed by Vinutha {\it et al.}~\cite{Vinutha2016natp}.
Even in the AQS protocols, the number of the strain step per cycle
$N_{\rm cycle}$ is finite and, therefore,  
the aperture between the pair of particles at contact, 
$r_{ij} -\sigma_{ij}$ where $\sigma_{ij}\equiv (\sigma_i + \sigma_j)/2$, 
is inevitably finite.
To remove this unwanted effect and obtain the true contact number in the $N_{\rm cycle}=\infty$ limit, 
we introduce a small `tolerant' inter-particle gap length, $a_{\rm th}$. 
With this parameter, the contact number $Z$ is reasonably estimated by 
\be
Z = \int_{a_{ij}}^{a_{ij}+a_{\rm th}} 2\pi r g(r) {\rm d}r.
\label{coord}
\ee
We calculated $Z$ as a function of $a_{\rm th}$ 
and assured that $Z$ saturates and reaches a limiting value in the limit $a_{\rm th}\to 0$.  
We fix $a_{\rm th}$ to be $a_{ij}/N_{\rm cycle}$ which is small enough for $Z$ to stay in this
limiting value~\cite{Vinutha2016natp}.   

\begin{figure}
\includegraphics[width=1.0\columnwidth,angle=-0]{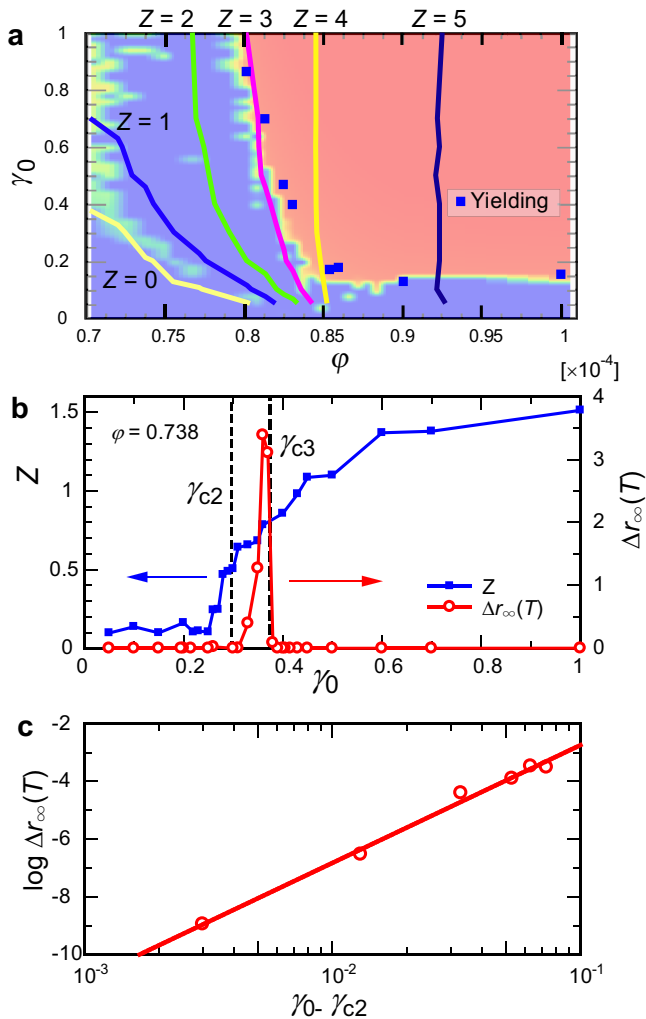}
\caption{\label{fig4}
(a):  The iso-$Z$ lines superimposed on the phase diagram shown in
 Fig.~\ref{fig2}{\bf a}.
They are obtained from the AQS protocol. 
 Square dots are the yielding transition points obtained from the steady shear protocol.
See Fig.~\ref{fig5} for more details. 
(b): $\gamma_0$-dependence of $Z$ and $\Delta r_{\infty}(T=\infty)$ at
 $\varphi=0.738$ obtained from the AQS protocol.  
(c): Replot of $\Delta r_{\infty} (T=\infty)$ shown in Fig.~\ref{fig4}{\bf b} as a function of 
$\gamma_0-\gamma_{\rm c2}$.
The solid line is the fit by $\Delta r_{\infty} (T=\infty)
= A(\gamma_0-\gamma_{\rm c2})^{\beta}$ with 
$\beta= 4.1$ and $\gamma_{\rm c2}= 0.29$. 
}
\end{figure}

In Figure~\ref{fig4}{\bf a}, the iso-$Z$ lines for several $Z$'s
obtained using the above-mentioned method are 
superimposed on top of the phase diagram shown in Figure~\ref{fig2}{\bf a}.  
One first observes the isostatic line with $Z=4$ ($=2d$) running
vertically at $\phi \approx  \phiJ$, 
as it is expected from the Maxwell's stability criterion~\cite{vanHecke2010}. 
A small deviation of the intersection at $\gamma_0=0$ 
from the true $\phiJ \approx 0.842$ is due to the small system size.
The line with $Z=5$ is located at much higher density and run almost
parallel with the isostatic line. 
The line with $Z=3$ starts slightly below $\phiJ$ at $\gamma_0=0$ but bends toward lower
$\phi$'s as $\gamma_0$ increases, and merges to the phase boundary 
of the discontinuous RI transition line $\gamma_{\rm c1}$ 
which eventually converges to $\phi \approx 0.80$,  the random loose packing density, 
This behavior is analogous to the finding for the three-dimensional system~\cite{Vinutha2016natp}, 
in which  the configurations obtained at $Z=d+1$ under shear are analogous to those of the jammed frictional
particles.
The line with $Z=2$ coincides with the iso-$\tau_{\rm L}$ line with $\tau_{\rm L}=1000T$
shown as the orange colored region in Fig.~\ref{fig2}{\bf c}. 
The line with $Z=1$ almost matches with the phase boundary $\gamma_{\rm c3}$ where the
discontinuous and reentrant RI transition takes place.
Finally, there is a region of $Z=0$  in small  $\phi$ and $\gamma_0$ region,
where the system is in the point-reversible state. 
As $\gamma_0$ is increased for a fixed $\phi$, there is a point at which
$Z$ becomes finite. 
Interestingly, these points agree well with the phase boundary
$\gamma_{\rm c2}$ where the continuous RI transition takes place. 
The yellow line in Figure~\ref{fig4}{bf a} represents the points at which $Z$ exceeds $0.2$. 

In Figure~\ref{fig4}{\bf b}, we shows the $\gamma_0$-dependence of both
$\Delta r_{\infty} (T=\infty)$ and $Z$ 
at a fixed $\varphi\approx 0.738$. 
Here $\Delta r_{\infty} (T=\infty)$ is evaluated using the AQS protocol and
with much long longer cycles of $n=40000$ than those used to draw
Fig.~\ref{fig2}{\bf a} to exclude unwanted metastable samples.   
One observes that $Z$ starts growing continuously from zero to finite values around 
the transition point $\gamma_{\rm c2}$ and reaches $Z=1$ in the vicinity of $\gamma_{\rm c3}$ where
$\Delta r_{\infty} (T=\infty)$ drops to zero discontinuously. 
This figure also shows clearly the continuous growth of $\Delta r_{\infty} (T=\infty)$ 
at $\gamma_{\rm c2}$. 
However, one realizes that its critical growth is qualitatively different from those observed
at very low densities where the RI transition is akin to those of the DP universality
class~\cite{Corte2008}. 
Figure~\ref{fig4}{\bf c} shows $\Delta r_{\infty} (T=\infty)$ as a function of
the distance from RI transition point $\gamma_0-\gamma_{\rm c2}$. 
The result is well fitted by 
\be
\Delta r_{\infty} (T=\infty)= A(\gamma_0-\gamma_{\rm c2})^{\beta}
\ee
with an exponent $\beta \approx 4.1$ and $\gamma_{\rm c2} \approx 0.29$. 
This exponent is far larger than those of the conserved DP universality class, or Manna class, 
$\beta_{\rm DP}=0.624\pm 0.029$~\cite{Lubeck2004ijmpb}. 
The discrepancy from the Manna class is not surprising because 
the transition at this density is escorted by an encoded geometrical 
change manifested by the increasing contact number and it is qualitatively
 distinct from the DP transition where the configuration of the particles is completely random.

\subsection{Yielding transition below $\phiJ$}
\begin{figure}
\includegraphics[width=0.95\columnwidth,angle=-0]{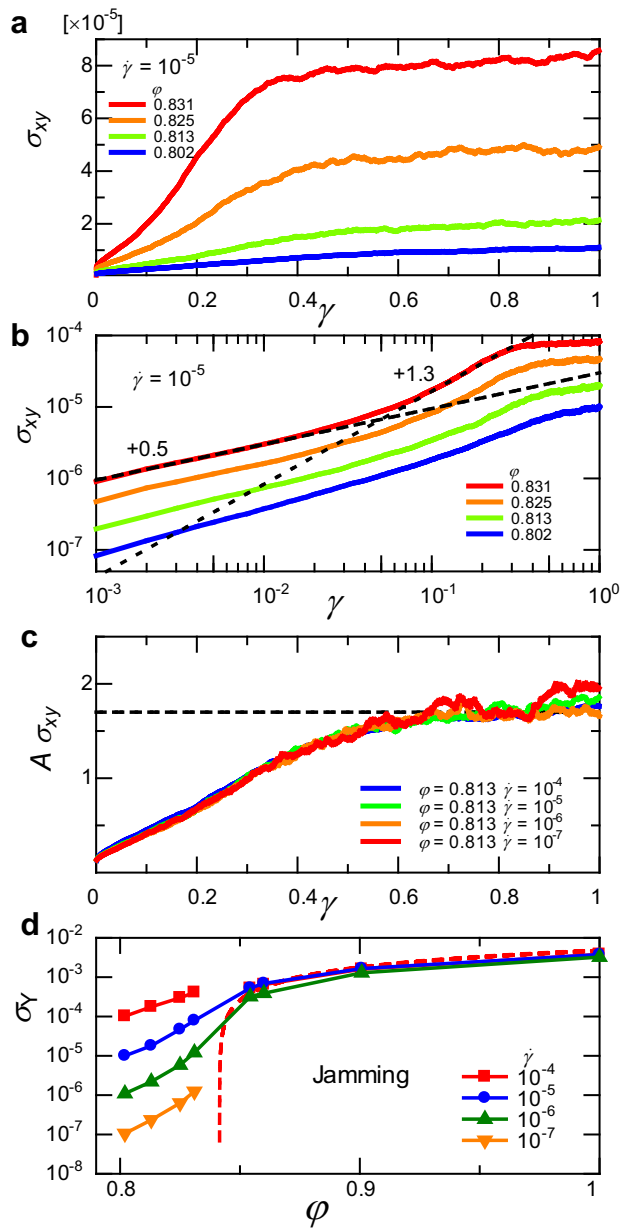}
\caption{\label{fig5} 
(a):  $\gamma$-dependence of the shear stress $\sigma_{xy}$ for various $\phi$
at finite shear rate $\gdot=10^{-5}$.
(b):  The same as (a) but at small $\gamma$'s in the log-log plot. 
Dotted and broken lines are the fit with $\gamma^{0.5}$ and $\gamma$, respectively.
(c):  $\gamma$-dependence of the scaled shear stress $A\sigma_{xy}$ at $\phi=0.813$, where 
$A$ is set to be $A\sim \gdot^{-1}$. The dash line represents the steady state of $\sigma_{xy}$ above shear yielding point$\gamma_{\rm Y}$. 
(d): $\phi$ dependence of yield stress $\sigma_{\rm Y}$ for various $\gdot$. The dash line represents  
$\sigma_{xy} = C(\phi-\phi_{\rm J})$, where $C=0.03$ and $\phi_{\rm J}=0.8425$.}
\end{figure}

Lastly, let us focus on the $\gamma_{\rm c1}$-line slightly below $\phiJ$.
At $\phi >\phiJ$,  $\gamma_{\rm c1}$ matches with the yielding transition line as a natural consequence that the elastic response of the system is associated with the reversible trajectories and 
the trajectories become diffusive in the plastic flow regime. 
It is somewhat surprising that the RI transition line still survives 
even below $\phiJ$ where 
the system is in the fluid phase and the yielding transition is
completely absent if one uses the AQS protocol. 
As we demonstrated in the previous subsection, $\gamma_{\rm c1}$ below
$\phiJ$ matches with the  iso-$Z$  line with $Z=3$. 
The presence of this RI transition slightly below $\phiJ$ is already reported by 
Schreck {\it et al.}~\cite{Schreck2013pre} and
Vinutha {\it et al.} have reported that the sheared frictionless
particles below $\phiJ$ develop structural configurations with 
$Z=d+1$ which would stabilize the system mechanically if the friction
force is turned on~\cite{Vinutha2016natp}. 
This implies that the information of the mechanical properties are 
embedded in the sheared configurations.
Such a hidden mechanical information would be detected either by
turning on the friction force between the particles or by observing the transient mechanical response
with time dependent strain.
We adopt the latter strategy, {\it i.e.}, 
we exert the shear strain with {\it finite} strain rate on the system.
We use the simple unidirectional shear with a constant strain rate
$\dot{\gamma}$ and the particles are driven with Eq.~(\ref{em}).  
We measure the shear stress $\sigma_{xy}$ defined by 
\begin{equation}
\sigma_{xy}
=\frac{1}{2L^2}\sum_{j,k}\frac{x_{jk}y_{jk}}{r_{jk}^2}\frac{\partial U}{\partial r_{jk}}. 
\end{equation}
Figure~\ref{fig5}{\bf a} is the strain ($\gamma$) dependence of $\sigma_{xy}$ 
at $\gdot=10^{-5}$ for various $\phi$ below $\phiJ$. 
The stress-strain curves are averaged over 1000 independent initial configurations 
which are generated by the AQS oscillatory shear simulations at
$\gamma_0=1.0$ for each packing fraction.  
These stress-strain curves demonstrate typical yielding behaviors from the elastic 
to the flowing phase; 
$\sigma_{xy}$ grows linearly with $\gamma$ and then yields at
$\gamma=\gamma_{\rm Y}$ followed by the steady plastic flow regime where
$\sigma_{xy}$ is constant.   
Interestingly, the yielding transition point $\gamma_{\rm Y}$ increases when $\phi$
decreases and it matches with $\gamma_{\rm c1}$ as shown in
Fig.~\ref{fig4}{\bf a}.  
This behavior is very similar to the that observed experimentally in
athermal colloidal suspensions below the glass transition point~\cite{Mason1996jcis}.

One observes the elastic regime before the yielding transition. 
However, the close inspection reveals that the elastic response against
$\gamma$ is not linear. 
Figure~\ref{fig5}{\bf b} is the log-log plot of the stress-strain curve
at very small $\gamma$ for several $\phi$'s, where 
the residual stress at $\gamma=0$ is excluded from $\sigma_{xy}$.
We find that $\sigma_{xy}$ follows the nonlinear 
behavior which is well fitted by
\begin{equation}
 \sigma_{xy} \propto  \sqrt{\gamma}.
\end{equation} 
This is reminiscent of the softening of the stress observed above
$\phiJ$~\cite{Otsuki2014pre,Nakayama2016jsmte,Dagois-Bohy2017sm}. 
In the studies above $\phiJ$, the softening is observed after the
linear elastic regime at 
even smaller $\gamma$ but we did not observe the
linear region. 
Figure~\ref{fig5}{\bf b} demonstrates that 
the softening region crosses over to the another nonlinear region with
$\sigma_{xy} \propto  \gamma^{a}$ with a constant 
$a \approx 1.3$. 
This is again analogous with the findings reported above 
$\phiJ$~\cite{Dagois-Bohy2017sm}. 
This crossover $\gamma$ is very sensitive to $\phi$ and increases
as $\phiJ-\phi$  increases. 
This result suggests that the instability of the trajectories which
controls the quasi-plastic event above $\phiJ$ also exists below
$\phiJ$. 

Figure~\ref{fig5}{\bf c} is $\sigma_{xy}$ for a fixed $\phi$ and for
several ${\dot{\gamma}}$'s. 
It demonstrates that the amplitude of $\sigma_{xy}$ is linear to $\dot{\gamma}$.  
All  data are collapsed to a single line when scaled by a constant $A$
proportional to $\dot{\gamma}$, which implies that $\gamma_{\rm Y}$ is
insensitive to $\dot{\gamma}$.
In Fig.~\ref{fig5}{\bf d}, we show the yield stress $\sigma_{\rm Y}\equiv
\sigma_{xy}(\gamma=\gamma_{\rm Y})$ as a function of $\varphi$ for 
several $\dot{\gamma}$. 
While the value above $\varphi_{\rm J}$ is
insensitive to $\dot{\gamma}$,  $\sigma_{\rm Y}$ increases almost
linearly with $\dot{\gamma}$ below $\phiJ$.

\section{Summary and conclusions} 
\label{sec:conclusions}

To summarize, we numerically studied the reversible-irreversible (RI)
transitions in particle trajectories for a wide range of densities below
and above the jamming transition density $\phiJ$.
Particular emphasis is placed on the densities below $\phiJ$, where  
we have observed very rich phases and mechanical behaviors. 

Well below $\phiJ$, there is a transition from the
point-reversible to the irreversible phase when $\gamma_0$ is increased
for a fixed density before it enters the loop-reversible state at higher $\gamma_0$'s. 
The transition from the point-reversible to the irreversible phases is a 
continuous transition, whereas the irreversible to loop-reversible
transition is discontinuous.  
Although the simulation windows are limited and 
it may sound difficult to judge whether they are the true
nonequilibrium transitions or a transient crossover.  
However, the microscopic trajectories inside the irreversible region becomes diffusive after a very long ballistic initial growth.
It is surprising to observe such reentrant transition because, 
for the particle trajectory to close a loop at the end of one or multiple cycle(s),
a minimal amount of strain amplitude would be necessary for 
the particle to be driven along the pathway encoded in the complex
geometrical landscape of the system.

The RI transition slightly below $\varphi_{\rm J}$ was also surprising. 
The transition is discontinuous as it is the case above $\phiJ$ 
but the order parameter $\Delta r_\infty(T)$ 
is much smaller than that above $\phiJ$ and 
$\gamma_{\rm c}$ sharply increases as $\varphi$ depart from $\varphi_{\rm J}$. 
The system is completely stress-free, or $\sigma_{xy}=0$, in the AQS
limit and thus the yielding transition is absent.
However, the mechanical properties are encoded in the sheared
configurations of the particles even below $\phiJ$.
We found that the stress becomes finite and the yielding behaviors
emerge as the strain rate $\dot{\gamma}$ increases. 
The yielding transition
points agree with the RI transition line irrespective of $\dot{\gamma}$.
We also show the elastic response at small strains is highly nonlinear 
and proportional to $\sqrt{\gamma}$ as it was recently found above
$\phiJ$~\cite{Otsuki2014pre,Nakayama2016jsmte,Dagois-Bohy2017sm}.  
This implies the jamming criticality which controls mechanical and 
rheological behaviors near $\phiJ$ survive below $\phiJ$.

All these rich nonequilibrium behaviors are encoded in the
geometrical properties of the configuration generated by shear. 
The contact number $Z$ is a good measure and we found that the iso-$Z$
lines with different $Z$ are placed on top of the phase boundaries of
the RI transition below $\phiJ$. 
The line at which $Z$ becomes finite matches with the phase boundary
between the point-reversible and irreversible phase ($\gamma_{\rm c2}$). 
$Z=1$ corresponds to $\gamma_{\rm c3}$ where the discontinuous transition from
the irreversible to loop-reversible phase takes place.
The line of $Z=2$ lies at the region where one observes the crossover 
of the dynamics of $\left\langle \Delta r(t)\right\rangle$ from the two-step
to single exponential relaxation. 
And the line of $Z=3$ matches with $\gamma_{\rm c1}$ at $\phi <\phiJ$,  where
the system would acquire the mechanical rigidity if the particles had the
frictions~\cite{Vinutha2016natp}. 
We emphasize that the densities which we explored here are still large
compared with the systems explored in
Refs.~\cite{Pine2005,Corte2008,Milz2013pre,Tjhung2015prl}
where the system is very dilute and the reversible trajectories are
always affine-like, or point-reversible. 
In contrary, 
percolations of the sheared particles' configurations and the multiple
collisions between particles are essential in the system we studied.  
Therefore, the continuous transition which is reported here is not a simple extrapolation 
of the RI phase transition line at the dilute limit and the nature of
the RI transition is also very different. 
Indeed, the critical exponent for $\Delta r$ at $\gamma_{\rm c2}$ (see
Fig.~\ref{fig4}{\bf c}) is distinct from those predicted for the
direct percolation theory and our system belongs to the
different universality class. 

The missing piece in this study is the nature of the RI transition just
at $\phiJ$, which we deliberately avoid studying because the interplay
of the jamming criticality and the RI transition makes the analysis of
the trajectories very challenging~\cite{Lavrentovich2017pre}.   
We only reported the overall behavior of the order parameters such as $\Delta r_\infty(T)$ 
(see Fig.~\ref{fig2}{\bf b}) which tends to vanish at $\phiJ$. 
However, it is premature to conclude that this is the compelling
evidence of the continuous transition because the resolution of the
parameters is still low to zero in the critical regime and there is no
sign of the relaxation time $\tau_{\rm R}$ (see Fig.~\ref{fig2}{\bf d}).
This is the region where the intricate microscopic trajectories affect the nontrivial and nonlinear
rheological response, especially like the softening observed in this
study below $\phiJ$ and in the literatures above $\phiJ$~\cite{Otsuki2014pre,Nakayama2016jsmte,Dagois-Bohy2017sm}. The studies in this direction are left for future work.
Another interesting issue is the memory effects which recently attract
much attentions in the context of the RI transitions in athermal particle
systems~\cite{Keim2011prl,Fiocco2014prl,Keim2019rmp}. 
It was revealed that ``the memory'' of particles' configurations can be
encoded in the systems by a cyclic training but the nature of the
imprinted memory  in the high density amorphous solids~\cite{Fiocco2014prl} (in the loop-reversible
phase) are very different from that for the dilutes systems (in the
point-reversible phase). 
The diverse phase diagram shown in this paper hints 
that rich memory effects are observed even below the jamming transition
point~\cite{Kawasaki2019}.

Finally, we remark that the rich dynamical transitions might not be
limited to the athermal particles systems but should be generally observed in
the hard condensed matter systems  in which the RI transitions are observed~\cite{Okuma2011prb,Brown2019njp,Mangan2018prl,Dobroka2017njp}. 
It would be interesting to study how the effect of the long-range interaction or the
absence of the clear jamming point in these systems influence the observed phase
diagrams.

\acknowledgments
We thank L. Berthier, S. Sastry, K. A. Takeuchi, H. Hayakawa, M. Otsuki, and E. Tjhung for valuable discussions. 
The research leading to these results has received funding
from JSPS Kakenhi 
(No. 25000002. 
H16H04034, 
15H06263, 
16H04025, 
16H06018,
18H01188,
and 
19K03767).


\end{document}